\documentclass[letterpaper,12pt,tightenlines,notitlepage,superscriptaddress,nofootinbib,showpacs]{revtex4-1}
\usepackage{amsmath,amssymb}
\usepackage[pdftex]{graphicx}
\usepackage{mathptmx}
\usepackage{color}
\usepackage{bm,wasysym,mathtools}
\usepackage[colorlinks]{hyperref}

\newcommand{\svector}[2]{\begin{pmatrix}#1 \\ #2 \end{pmatrix}}

\begin{document}

\title{Speedmeter scheme for gravitational-wave detectors based on EPR quantum entanglement}

% \date{\today}

\author{E.~Knyazev}
\affiliation{M.V.Lomonosov Moscow State University, Faculty of Physics, Moscow 119991, Russia}

\author{S.~Danilishin}
\affiliation{SUPA, University of Glasgow, Glasgow G12 8QQ, United Kingdom}

\author{S.~Hild}
\affiliation{SUPA, University of Glasgow, Glasgow G12 8QQ, United Kingdom}

\author{F.Ya.~Khalili}
\email[]{Corresponding author: khalili@phys.msu.ru}
\affiliation{M.V.Lomonosov Moscow State University, Faculty of Physics, Moscow 119991, Russia}

\begin{abstract}
  We propose a new implementation of a quantum speed meter QND measurement scheme. It employs two independent optical readouts of the interferometer test masses, featuring  strongly different values of the bandwidths $\gamma_{1,2}$ and of the optical circulating power $I_{1,2}$, with the special relationship of $I_1/I_2=\gamma_1/\gamma_2$. The  outputs of these two position meters have to be combined by an additional beamsplitter. In this scheme, signals at the common and the differential outputs of the interferometer setup are proportional to the position and the velocity of the test masses, respectively. The influence of the position meter-like back action force associated with the position signal can be cancelled using the EPR approach by measuring the amplitude quadrature of the beamsplitter common output correlated with this force. In the standard signal-recycled Michelson interferometer topology of the modern gravitational-wave detectors, two independent optical position meters can be implemented by two orthogonal polarisations of the probe light. Our analysis shows that the EPR speedmeter provides significantly improved sensitivity for all frequencies below $\sim 30\,{\rm Hz}$ compared  to  an equivalent signal recycled Michelson interferometer. We believe the EPR speedmeter scheme to be very attractive for future upgrades of gravitational wave detectors, because it requires only minor changes to be implemented in the interferometer hardware and allows to switch between the position meter and the speed meter modes within short time-scales and without any changes to the hardware.
\end{abstract}

\maketitle

\section{Introduction}

The sensitivity of the modern laser-interferometric gravitational-wave (GW) detectors  is limited by quantum fluctuations of the probing light over most of the sensitive frequency range. In particular, at higher frequencies their sensitivity is limited by the shot noise  (also known in more general context as the measurement noise), created by quantum fluctuations of the phase of the probing light \cite{CQG_32_7_074001_2015, Affeldt_CQG_31_224002_2014, Accadia2012}. The resulting sensitivity, about $\sim10^{-20}\,{\rm m/\sqrt{Hz}}$ in units of the equivalent displacement noise, is extremely high and  has proved to be sufficient for the direct observation of gravitational waves from astrophysical sources \cite{PRL_116_131102_2016, PRL_116_241103_2016}. 

At the same time the pair of Advanced LIGO interferometers, which detected the first GW signals, have not reached yet their design sensitivity, which is planned to provide about a factor three improvement in astrophysical reach \cite{PRL_116_131103_2016}. Suppression of the shot noise, which is necessary for achieving this goal, will require either an increase of the optical power circulating in the interferometer up to $\sim1\,{\rm MW}$, or the application of squeezed light states \cite{Caves1981, Nature_2011, Nature_2013}, and most probably a combination of both approaches will be used to maximise the sensitivity gain.

Due to the Heisenberg uncertainty relation, this will lead to the proportional increase of another kind of the quantum noise, namely radiation pressure noise (also known as the quantum back action noise), imposed by the quantum fluctuations of the light power in the interferometer disturbing the test mass positions. The point of balance between the measurement noise and the back action noise is known as the Standard Quantum Limit (SQL) \cite{92BookBrKh}, and the design sensitivity of the Advanced LIGO interferometers will touch the SQL at one frequency. 

It has to be emphasized that the SQL is not a truly fundamental limit, and several methods have been proposed for overcoming the SQL in future  GW detectors. A detailed review of these methods can be found {\it e.g.} in \cite{12a1DaKh}. One of the most promising approaches for surpassing the SQL is based on the {\it quantum speed meter} concept, which was first proposed in \cite{90a1BrKh}. The basic idea of this concept is to measure the velocity of the probe mass(es) instead of their position. In this case, the measurement noise and the back action noise spectral densities depend on the observation frequency in such a way that  they can provide cancellation of each other by means of introducing a frequency-independent cross-correlation between them. It can be implemented simply by using a homodyne detector with the properly set homodyne angle. Note that in the traditional position-sensitive interferometers, additional long {\it filter cavities} are required for this type of the quantum noises cancellation \cite{02a1KiLeMaThVy} (4-km cavities were proposed in \cite{02a1KiLeMaThVy}; it was shown later that much shorter, but still quite long, tens or hundreds of meters, cavities could be used as well, but they could provide only limited sensitivity gain \cite{06a2Kh, 10a1Kh, Evans_PRD_88_022002_2013}).

Several implementations of the quantum speed meter concept suitable for the GW detectors were proposed, which can be divided into the following two categories: the first one relies on the ordinary Michelson interferometer topology of the contemporary GW detectors, but requires an additional long {\it sloshing cavity} \cite{00a1BrGoKhTh, Purdue2002} and therefore does not provide significant advantages in comparison with the filter cavities based topologies. The second category is based on the zero-area Sagnac interferometer topology \cite{Chen2002, 04a1Da}, which significantly deviates from the standard Michelson topology. Currently it is a subject of intense R\&D efforts \cite{Graef_CQG_31_215009_2014, Danilishin_NJP_17_4_043031_2015, Leavey_1603_07756, Houston_speedmeter2017}.

Here we propose a new kind of the quantum speed meter, the {\it EPR speed meter} (from the famous gedanken experiment by Einstein, Podolsky and Rosen), which allows to use the Michelson interferometer topology and, at the same time, does not require any additional long-baseline filter cavities or other major infrastructure changes. 

We would like to emphasize that the goal of this short paper is to introduce the concept of this new speed meter type. Detailed investigations of the technical implementation, such as the robustness against optical loss, the coupling of laser frequency and amplitude noise, as well as additional add-on techniques, like the injection of squeezed states will be considered in a follow-up article, currently in preparation.

This paper is organized as follows. In the next section we reproduce the basic analytical treatment of  quantum noise in the position meter and speed meter schemes. In Sec.\,\ref{sec:idea} we present the concept of the EPR speed meter. In Sec.\,\ref{sec:GWdetector} we consider a possible implementation of our concept in a GW wave detector and provide brief estimates of its sensitivity, using parameters similar to the ones of the envisaged  LIGO Voyager GW detectors \cite{wp2016}. The notations and the parameter values used in this paper are listed in Table\,\ref{tab:params}. 

\begin{table}
  \begin{ruledtabular}
    \begin{tabular}{ll}
      Quantity & Description \\ \hline
      $c$                 & Speed of light \\
      $\hbar$             & Reduced Plank constants \\
      $M = 200\,{\rm kg}$ & Reduced mass of the interferometer equal to the mass \\
      & of each of the arm cavities mirrors \cite{12a1DaKh} \\
      $L = 4\,{\rm km}$   & Length of the interferometer arm cavities \\
      $\omega_o = 2\pi c/1.550\,\mu{\rm m}$ & Resonance frequency of the interferometer and the optical pump frequency \\
      $I_c = 2\times3\,{\rm MW}$ & Total optical power circulating in the both arms of the interferometer \\
      $J = \dfrac{4\omega_oI_c}{MLc} = (2\pi\times79\,{\rm Hz})^3$ & Normalized optical power in the interferometer\\
      $\gamma$            & Half-bandwidth of the interferometer \\
      $\Omega$            & Audio sideband frequency of the GW signal \\
      $\zeta$             & Homodyne angle \\
%       $e^r$               & Squeezing factor \\
%       $\theta$            & Squeezing angle \\
%       $\eta$            & Unified quantum efficiency
    \end{tabular}
  \end{ruledtabular}
  \caption{Main notations used in this paper. For the numerical values, we use the ones planned for the next-generation GW detector LIGO Voyager \cite{wp2016}.}\label{tab:params}
\end{table}

\section{General  introduction to  quantum noise of the position meter and the speed meter}\label{sec:sm_vs_pm} 

\subsection{Position meter}

The (double-sided) power spectral density of the sum of quantum noise components in a position meter can be presented as follows  (a much more detailed analysis of the quantum noise in interferometers can be found in \cite{12a1DaKh}):
\begin{equation}\label{S_PM} 
  S_{\rm PM} = S_{xx} - \frac{2S_{xF}}{M\Omega^2} + \frac{S_{FF}}{M^2\Omega^4}\,,
\end{equation} 
where $S_{xx}$ is the spectral density of the measurement noise, $S_{FF}$ is the spectral density of the back action force and $S_{xF}$ is the cross-correlation spectral density of these two noise sources (we assume here that $S_{xF}$ is real in order to avoid subtle but unrelevant to our consideration issues related to the imaginary part of $S_{xF}$). These spectral densities satisfy the Heisenberg uncertainty relation
\begin{equation}\label{Heisenberg} 
  S_{xx}S_{FF} - S_{xF}^2 \ge \frac{\hbar^2}{4} \,.
\end{equation} 
In the rest of this section, we assume this relation is saturated and the interferometer is driven by the vacuum and laser fields in the minimum uncertainty quantum state.

In the modern GW detectors, there is no cross-correlation between the shot noise and the radiation pressure noise, because the resonance-tuned configuration is used in these detectors and only the phase quadrature of the outgoing light is measured. Hence
\begin{equation}\label{S_PM_SQL} 
  S_{\rm PM} = S_{xx} + \frac{S_{FF}}{M^2\Omega^4} \ge S_{\rm SQL} \,,
\end{equation} 
where
\begin{equation}\label{S_SQL} 
   S_{\rm SQL} = \frac{\hbar}{M\Omega^2} 
\end{equation} 
is the double-sided SQL spectral density.

On the other hand, if $S_{xF}\ne0$ and can be made arbitrarily dependent on frequency, then the spectral density \eqref{S_PM} can be minimized, using the exact equality in \eqref{Heisenberg} and setting
\begin{equation}\label{var_meas} 
  S_{xx} = \frac{\hbar^2}{4S_{FF}} + \frac{S_{FF}}{M^2\Omega^4} \,,\qquad 
  S_{xF} = \frac{S_{FF}}{M\Omega^2} \,,
\end{equation} 
which gives:
\begin{equation}\label{S_PM_opt} 
  S_{\rm PM} = \frac{\hbar^2}{4S_{FF}} \,.
\end{equation}
In the laser interferometers $S_{FF}$ is proportional to the optical power inside the interferometer. It was shown in \cite{00p1BrGoKhTh} that this {\it Energetic Quantum Limit} actually is a general one for all linear stationary interferometric measurements.

The optimized spectral density \eqref{S_PM_opt}, in principle, can be made arbitrarily small simply by increasing this power. However, conditions \eqref{var_meas} can only be satisfied in the given frequency band, provided that the spectral densities $S_{xx}$, $S_{xF}$, and $S_{FF}$ depend on frequency $\Omega$ in a rather specific way which is, sadly, different from the one they acquire due to finite bandwidth of the arm cavities in the existing GW interferometers. Therefore, to introduce the desired frequency dependence \eqref{var_meas} in a broad band, long additional filter cavities are required \cite{02a1KiLeMaThVy}.

\subsection{Speed meter}

In the speed meter schemes, the quantum noise has the same general structure \eqref{S_PM}, but with the following peculiarities:
\begin{equation}
  S_{xx} = \frac{S_{vv}}{\Omega^2}\,,\qquad S_{FF} = \Omega^2S_{pp}\,,\qquad 
  S_{xF} = -S_{vp} \,,
\end{equation} 
where $S_{vv}$ is the velocity measurement noise spectral density, $S_{pp}$ is the momentum perturbation noise spectral density, and $S_{vp}$ is the corresponding cross-correlation spectral density. It is important that $S_{vv}$, $S_{pp}$, and $S_{vp}$ can be considered as frequency-independent within the interferometer bandwidth \cite{00a1BrGoKhTh}. The relation \eqref{Heisenberg} takes the following form:
\begin{equation}\label{Heisenberg_v} 
  S_{vv}S_{pp} - S_{vp}^2 = \frac{\hbar^2}{4} \,,
\end{equation} 
and the sum quantum noise spectral density of the speed meter reads:
\begin{equation}\label{S_SM} 
  S_{\rm SM}
  = \frac{1}{\Omega^2}\left(S_{vv} + \frac{2S_{vp}}{M} + \frac{S_{pp}}{M^2}\right) .
\end{equation} 
In the particular case of $S_{vp}=0$, similar optimization as for the PM can be made:
\begin{equation}\label{opt_SM_SQL}
  S_{vv} = \frac{\hbar^2}{4S_{pp}} = \frac{\hbar}{2M} \,,
\end{equation} 
yielding the quantum noise of the speed meter to follow the SQL:
\begin{equation} 
  S_{\rm SM} = \frac{\hbar}{M\Omega^2} \,.
\end{equation} 
Note that the corresponding spectral density of the position meter \eqref{S_PM_SQL} only touches the SQL at one given frequency and goes above it elsewhere. Therefore, the speed meter provides better sensitivity even in this simple case. 

In a more general case of $S_{vp}\ne0$, the following optimization:
\begin{equation}\label{opt_SM} 
  S_{vv} = \frac{\hbar^2}{4S_{pp}} + \frac{S_{pp}}{M^2} \,,\qquad 
  S_{vp} = -\frac{S_{pp}}{M} \,,
\end{equation} 
gives:
\begin{equation}\label{S_SM_opt} 
  S_{\rm SM} = \frac{\hbar^2}{4\Omega^2S_{pp}} \,.
\end{equation} 
Similar to the position meter case \eqref{S_PM_opt}, this spectral density can be arbitrary small, provided that $S_{pp}$ is sufficiently large, which means high enough circulating optical power in the interferometer. Contrary to the position meter, no additional elements like filter cavities are required for this.

\section{Idea of the EPR speed meter}\label{sec:idea} 

\begin{figure}
    \includegraphics[width=0.3\textwidth]{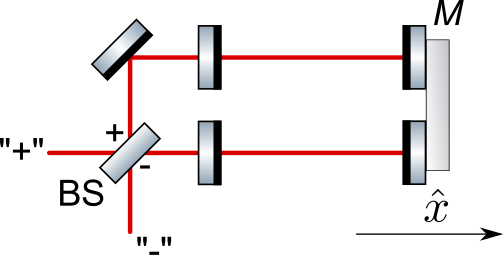} \\[1ex]
    \includegraphics[width=0.42\textwidth]{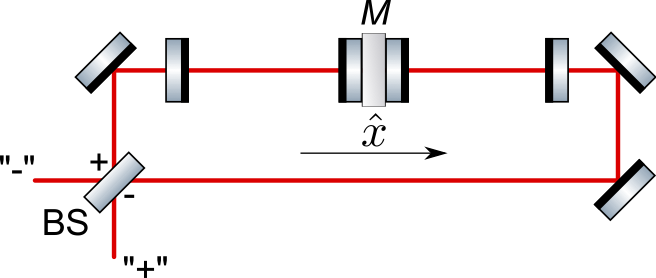}   
  \caption{Conceptual schemes of the EPR speed meter. \textit{Top}: two optically independent Fabry-Perot cavities sense the position $x$ of the same mass $M$; their output beams are combined by the beamsplitter, forming the ``$+$'' (position) and the ``$-$'' (speed) outputs. Bottom: a more practical collinear version (tolerant to the angular motion of the mass $M$) of the same scheme; in this case, the signs of the beamsplitter reflectivity factors has to be swapped.}\label{fig:sketch_scheme} 
\end{figure} 

Consider now the scheme shown in Fig.\,\ref{fig:sketch_scheme}(top). Here the mass $M$ forms a joint movable mirror for two otherwise independent Fabri-Perot cavities having the same eigenfrequency $\omega_o$, the same lengths $L$, but different bandwidths $\gamma_{1,2}$. The cavities are pumped at the frequency $\omega_o$ and their output fields are combined by the beamsplitter. Its two output beams labeled in the picture as ``$+$'' and ``$-$''are measured by the two homodyne detectors. 

Using the two-photon amplitudes notations of \cite{Caves1985, Schumaker1985}, the input/output relations for these cavities can be written as (see {\it e.g.} \cite{16a1DaKh}):
\begin{equation}
  \svector{\hat{{\rm b}}^c_{\rm 1,2}}{\hat{{\rm b}}^s_{\rm 1,2}}
  = \mathcal{R}_j\svector{\hat{{\rm a}}^c_{\rm 1,2}}{\hat{{\rm a}}^s_{\rm 1,2}}
    + \mathcal{G}_j\svector{0}{\hat{x}} ,
\end{equation} 
where $j=1,2$ is the cavity number, $\hat{{\rm a}}_j^{c,s}$ are the cosine and the sine quadratures of the input field of the cavity $j$, $\hat{{\rm b}}_j^{c,s}$ are the corresponding output field quadratures, 
\begin{equation}
  \mathcal{R}_j = \frac{\gamma_j+i\Omega}{\gamma_j-i\Omega}
\end{equation} 
are the frequency-dependent reflectivities of the cavities for the cavity sideband fields,
\begin{equation}
  \mathcal{G}_j 
  = \frac{2\sqrt{2}\omega_o{\rm E}_j}{\gamma_j-i\Omega}\sqrt{\frac{\gamma_j}{cL}} 
\end{equation} 
are the optomechanical transfer functions, ${\rm E}_j$ are the classical amplitudes of the intracavity fields, normalized as follows:
\begin{equation}
  \hbar\omega_o{\rm E}_j^2 = I_j \,,
\end{equation} 
and $I_j$ is the optical power, circulating in the cavity $j=1,2$. Note that if $\hat{{\rm a}}_j^{c,s}$ correspond to the vacuum input fields, then the same is true for $\mathcal{R}_j\hat{{\rm a}}_j^{c,s}$. Therefore, below we absorb $\mathcal{R}_j$ into $\hat{{\rm a}}_j^{c,s}$ in order to simplify the equations.

The beamsplitter transforms the output fields as follows:
\begin{equation}\label{io_pm} 
  \svector{\hat{{\rm b}}^c_\pm}{\hat{{\rm b}}^s_\pm}
  = \svector{\hat{{\rm a}}^c_\pm}{\hat{{\rm a}}^s_\pm} 
    + \mathcal{G}_\pm\svector{0}{\hat{x}} ,
\end{equation} 
where 
\begin{equation}
  \hat{{\rm a}}^{c,s}_\pm 
  = \frac{\hat{{\rm a}}^{c,s}_1 \pm \hat{{\rm a}}^{c,s}_2}{\sqrt{2}}
\end{equation} 
are the new effective input vacuum fields and
\begin{equation}
  \mathcal{G}_\pm = \frac{\mathcal{G}_1 \pm \mathcal{G}_2}{\sqrt{2}} \,.
\end{equation} 
are the transfer functions for the ``$+$'' and ``$-$'' channels.

In order to create the speed meter type frequency dependence of the optomechanical coupling, we propose to exploit the difference in the frequency dependence of $\mathcal{G}_{1,2}$. Note that if $\gamma_1\ne\gamma_2$ and
\begin{equation}\label{the_cond} 
  \frac{{\rm E}_1}{\sqrt{\gamma}_1} = \frac{{\rm E}_2}{\sqrt{\gamma}_2} \,,
\end{equation} 
then the DC values of $\mathcal{G}_1$ and $\mathcal{G}_2$ are equal to each other, but the AC values are different. In this case, the leading term of the factor $\mathcal{G}_-$ is proportional to $\Omega$, as it is required for the speed measurement. More specifically, in this case
\begin{subequations}\label{calG_pm} 
  \begin{gather}
    \mathcal{G}_+ = \frac{2\omega_o{\rm E}_1}{\sqrt{\gamma_1cL}}\,
      \frac{2\gamma_1\gamma_2-i\Omega(\gamma_1+\gamma_2)}
        {(\gamma_1-i\Omega)(\gamma_2-i\Omega)} \,, \\
    \mathcal{G}_- = \frac{2\omega_o{\rm E}_1}{\sqrt{\gamma_1cL}}\,
      \frac{i\Omega(\gamma_2-\gamma_1)}{(\gamma_1-i\Omega)(\gamma_2-i\Omega)} \,.
      \label{calG_m}
  \end{gather}
\end{subequations}
Taking also into account that the ``$+$'' and ``$-$'' noise components in \eqref{io_pm} are uncorrelated, it follows from Eqs.\,\eqref{calG_pm} that the ``$+$'' and ``$-$'' ports of the beamsplitter correspond to two independent meters: the position one for the ``$+$'' port (note that $\mathcal{G}_+$ is flat at low frequencies) and the speed meter for the ``$-$'' port.

According to the Heisenberg uncertainty relation, each of these meters create the back action force of its own. Indeed, it is easy to show that the fluctuational component of the radiation pressure force acting on the mass $M$ is equal to (see again \cite{16a1DaKh}):
\begin{equation}\label{F_fl_sum} 
  \hat{F}_{\rm fl} 
  = \hbar(\mathcal{G}_1^*\hat{{\rm a}}_1^c + \mathcal{G}_2^*\hat{{\rm a}}_2^c)
  = \hat{F}_{\rm fl+} + \hat{F}_{\rm fl-} \,,
\end{equation} 
where
\begin{equation}\label{F_fl_pm} 
  \hat{F}_{\rm fl\pm} = \hbar\mathcal{G}_\pm^*\hat{{\rm a}}_\pm^c \,.
\end{equation} 
It is easy to see that the force $\hat{F}_{\rm fl-}$ has the speed meter type frequency dependence.

In principle, a more practical version with the collinear placement of the elements, which is tolerant to the angular motion of the test mass, is also possible, see Fig.\,\ref{fig:sketch_scheme}(bottom). Evidently, in this case the signs of reflectivities of the two beamsplitter surfaces has to be swapped. In all other aspects, the features of this version are identical to the one considered before.

In general the two outputs of the proposed two-meters setup can be combined in a variety of ways. Here we mainly focus at the simplest pure speed meter case, which can be implemented by ``switching off'' the position-sensitive ``$+$'' channel. This can be done by measuring the amplitude (cosine) quadrature of the ``$+$'' output field, which does not contain any information on the mechanical motion, but instead it is correlated with radiation pressure force $\hat{F}_{\rm fl+}$. Therefore,  it is entangled with the mechanical motion. The measurement of $\hat{{\rm a}}_+^c$ removes this entanglement in the EPR way and projects the mass $M$ into the conditional state with effectively eliminated perturbation component created by the ``$+$'' channel. Of course, physically this perturbation remains unchanged, but it becomes known to the experimenter and can be eliminated from the output signal by the subsequent appropriate data processing. It could be mentioned that in general case, this filtering can not be done by in real time by using a causal filter. However, for the GW detection this is not an issue because off-line cancellation with an acausal filter can be used in this case.

The remaining ``$-$'' channel quantum noise spectral density can be calculated using Eqs.\,(\ref{S_PM}, \ref{io_pm}, \ref{calG_m}, \ref{F_fl_pm}). Assuming that the output of this channel is measured by the homodyne detector with the homodyne angle $\zeta_-$, we obtain:
\begin{equation}\label{S_EPR} 
  S_{\rm EPR} = \frac{S_{\rm SQL}}{2}\left(
      \frac{1}{\mathcal{K}_-\sin^2\zeta_-} - 2\cot\zeta_- 
      + \mathcal{K}_-
    \right) ,
\end{equation} 
where 
\begin{equation}\label{K_EPR} 
  \mathcal{K}_- = S_{\rm SQL}|\mathcal{G}_-|^2
  = \frac{(\gamma_1-\gamma_2)^2J}
      {(\gamma_1+\gamma_2)(\gamma_1^2+\Omega^2)(\gamma_2^2+\Omega^2)} 
\end{equation} 
is the Kimble factor \cite{02a1KiLeMaThVy} for the ``$-$'' channel. The three terms in  parentheses in \eqref{S_EPR} originate, respectively, from the shot noise, the cross-correlation between the the shot noise and the radiation pressure noise, and the radiation pressure noise. They directly relate to the corresponding terms in Eqs.\,\eqref{S_PM} and \eqref{S_SM}.

It is instructive to compare $\mathcal{K}_-$ with the corresponding factors for the position meter (Michelson) interferometers \cite{02a1KiLeMaThVy}:
\begin{equation}\label{K_PM} 
  \mathcal{K}_{\rm PM} = S_{\rm SQL}|\mathcal{G}_{\rm PM}|^2 
  = \frac{2\gamma J}{\Omega^2(\gamma^2+\Omega^2)} 
\end{equation} 
and for the speed meter Sagnac interferometer \cite{Chen2002}:
\begin{equation}\label{K_SM}
  \mathcal{K}_{\rm SM} = S_{\rm SQL}|\mathcal{G}_{\rm SM}|^2 
  = \frac{4\gamma J}{(\gamma^2+\Omega^2)^2} \,.
\end{equation} 
Note that in order to fulfill the speed meter conditions (\ref{opt_SM_SQL}, \ref{opt_SM}), the optomechanical coupling has to be sufficiently strong, $\mathcal{K}_{\rm SM} \ge 1$, which requires sufficiently high circulating optical power:
\begin{equation}\label{J_Sagnac} 
  J \ge \frac{\gamma^3}{4} \,.
\end{equation} 

More sophisticated frequency dependence of $\mathcal{K}_-$ allows to alleviate this problem. Indeed $\mathcal{K}_-$ has a flat speed meter-like behaviour only up to the smaller of the two bandwidths $\gamma_{1,2}$; to be specific, we suppose that $\gamma_2<\gamma_1$. At higher frequencies $\Omega\gg\gamma_2$, it can be approximated as 
\begin{equation}\label{K_EPR_hf} 
  \mathcal{K}_- 
  \approx\frac{(\gamma_1-\gamma_2)^2J}{(\gamma_1+\gamma_2)(\gamma_1^2+\Omega^2)\Omega^2}
\end{equation} 
and has the position meter type frequency dependence with the bandwidth equal to $\gamma_1$, compare with \eqref{K_PM}. 

Consider therefore the strongly asymmetric case with $\gamma_2\ll\gamma_1$ and with $\gamma_1$ is equal to the required bandwidth of the interferometer. In this case the speed meter conditions (\ref{opt_SM_SQL}, \ref{opt_SM}) give the following requirement for the circulating optical power:
\begin{equation}
  J \ge \frac{(\gamma_1+\gamma_2)\gamma_1^2\gamma_2^2}{(\gamma_1-\gamma_2)^2} 
  \approx \gamma_1\gamma_2^2 
\end{equation} 
which is a factor of $1/4\times(\gamma_1/\gamma_2)^2$ weaker than \eqref{J_Sagnac}. With the account for this moderate optical power, the quantum noise at frequencies above the $\gamma_2$ will be dominated by the shot noise, with the position meter-like $\mathcal{K}_-$ providing $1/4\times(\gamma_1/\Omega)^2$ lower shot noise spectral density within the interferometer bandwidth ($\Omega\lesssim\gamma_1$) than in the Sagnac  speed meter case. 

The price to pay, however, is the approximately two times lower high-frequency asymptotic of $\mathcal{K}_-$ \eqref{K_EPR_hf} compared to that of the ordinary position meter \eqref{K_PM} (see also Fig.\,\ref{fig:plots_G}). The reason for this is evident: only the part of the total circulating optical power which corresponds to the ``$-$'' channel, that is about half of the total power, participates in the measurement. Whether this disadvantage outweighs the advantage of more effective $\mathcal{K}_-$ discussed above or not, depends on the specific parameters of the interferometer.

\section{Prospects of use in GW detectors}\label{sec:GWdetector} 

\begin{figure}[h]
% 	\centering
	\includegraphics[width=0.45\textwidth]{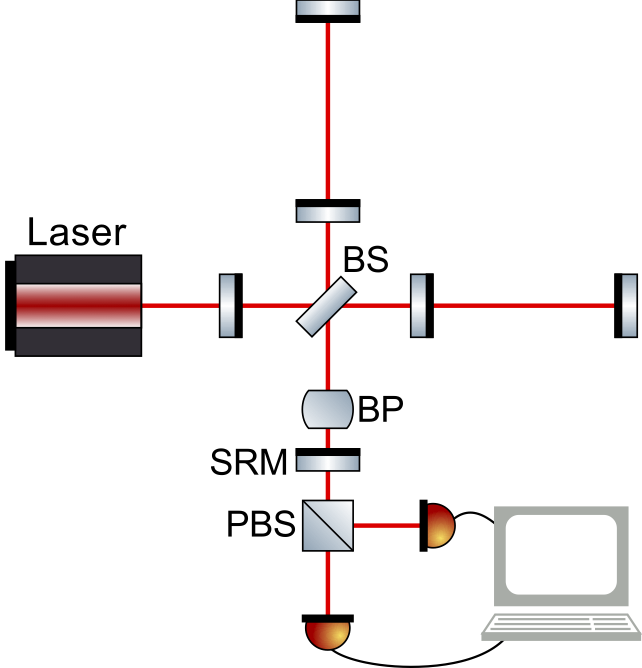}
	\caption{Possible implementation of the EPR speed meter based on the standard signal recycled interferometer topology of the modern GW detectors. SRM: signal recycling mirror; BP: birefringent plate; PBS: polarising beamsplitter.}	
	\label{fig:setup}
\end{figure}

A possible method of implementation of two independent position meters within the standard Michelson interferometer topology is to use the two orthogonal polarisations of light. Note that polarisation-based schemes of speed meter (the Sagnac interferometer or the Michelson interferometer with the additional sloshing cavity) were discussed already in literature several times \cite{Beyersdorf_JOSAB_16_1354_1999, Beyersdorf_OptLett_24_1112_1999, 02a2Kh, 04a1Da, Wade_PRD_86_062001_2012, Wang_PRD_87_096008_2013}. 

The scheme which we propose here is shown in Fig.\,\ref{fig:setup}. In this scheme, in order to excite two orthogonal linear polarisations of the carrier light in the interferometer, the polarisation of the pumping laser has to be tilted by the angle defined by the condition \eqref{the_cond}:
\begin{equation}
  \arctan\frac{E_2^2}{E_1^2} = \arctan\frac{\gamma_2}{\gamma_1}
\end{equation} 

Two strongly different values of the bandwidths $\gamma_1\gg\gamma_2$ can be created by using the signal recycled configuration of the interferometer \cite{Meers1988, Mizuno_PLA_175_273_1993} with a birefringent plate inserted into the signal recycling cavity. This plate has to introduce the phase shift $\pi/2$ between the two polarisations passing through it (the quarterwave plate). In this case, for one polarisation we obtain a resonant sideband extraction scheme (large bandwidth), while for the other polarisation we obtain a signal recycling scheme (low bandwidth). The bandwidths of the two readouts are given by:
\begin{subequations}
  \begin{gather}
    \gamma_1 = \frac{c}{4L}\,
      \frac{T_{\rm SRM}T_{\rm ITM}}{1-\sqrt{R_{\rm SRM}R_{\rm ITM}}}\,,\\
    \gamma_2 = \frac{c}{4L}\,
      \frac{T_{\rm SRM}T_{\rm ITM}}{1+\sqrt{R_{\rm SRM}R_{\rm ITM}}}\,,
  \end{gather}
\end{subequations}
where $R_{\rm ITM}$, $T_{\rm ITM}$ are the power reflectivity and transmissivity of the arm cavities input mirrors, and $R_{\rm SRM}$, $T_{\rm SRM}$ are the power reflectivity and transmissivity of the signal recycling mirror. 

Finally, in order to create the output channels ``$+$'' and ``$-$'', the polarisation beamsplitter tilted by $45^\circ$ relative to the linear polarisations circulating in the interferometer can be used. 

\begin{figure}
  \includegraphics[scale=1.2]{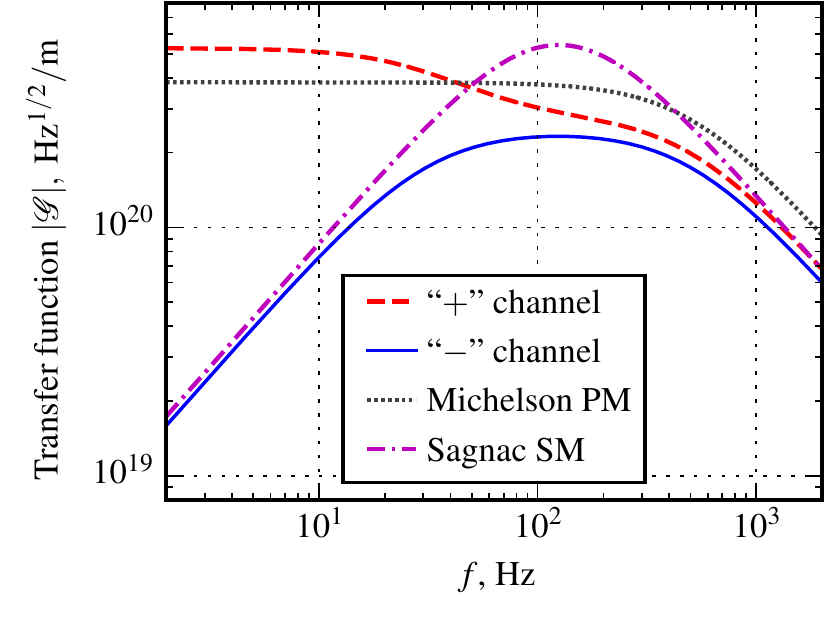}
  \caption{Optomechanical transfer functions for: the  ``$+$'' and the ``$-$'' channels of the EPR speedmeter noise with $\gamma_1=2\pi\times500\,{\rm Hz}$ and $\gamma_2=\sqrt{J/\gamma_1}\approx2\pi\times31\,{\rm Hz}$; the Michelson/Fabry-Perot position meter interferometer with $\gamma=2\pi\times500\,{\rm Hz}$; the Sagnac speed meter interferometer with $\gamma=(4J)^{1/3}\approx2\pi\times125\,{\rm Hz}$. All other parameters correspond to the ones of the planned LIGO Voyager GW detector, see Table \ref{tab:params}.}\label{fig:plots_G} 
\end{figure} 

For the sensitivity estimates for this scheme, we use the parameters planned for the next generation GW detector LIGO Voyager \cite{wp2016}, see Table \ref{tab:params}. Note that due to the increased mirror mass and the changed laser wavelength (compared to Advanced  LIGO) and despite the increased circulating optical power, this detector will be even more underpowered than the Advanced LIGO, with the normalized optical power $J\ll\gamma^3$. Therefore, we assume here in most cases that the phase quadrature of the output channel ``$-$'' is measured, which corresponds to homodyne angle $\zeta_-=\pi/2$, see Eq.\,\eqref{S_EPR}. In this regime, the speed meter can not overcome the SQL, but still has much better low-frequency sensitivity than the position meter, see Sec.\,\ref{sec:sm_vs_pm}. At the same time, this regime requires smaller optical circulating power than the more advanced SQL-beating one with $\zeta\ne\pi/2$. 

\begin{figure}
  \includegraphics[scale=1.2]{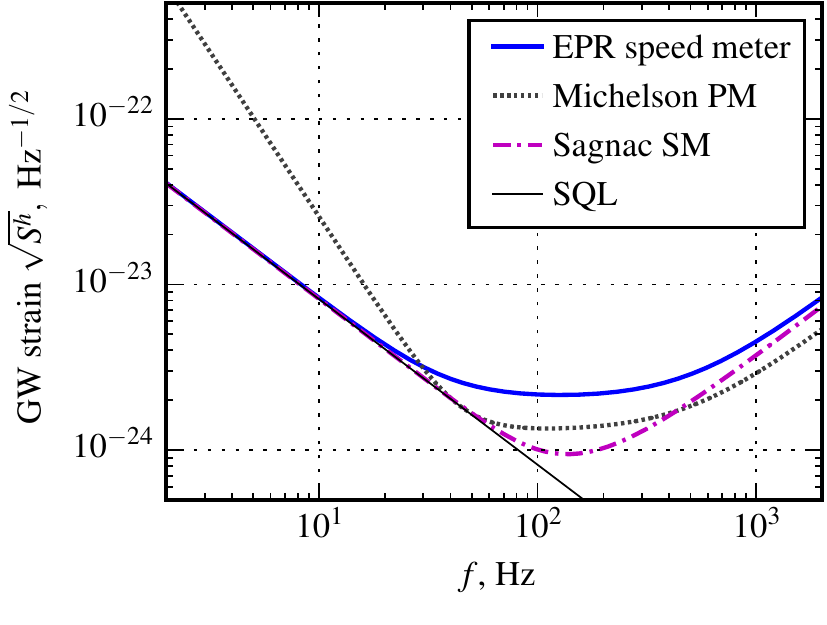}
  \caption{Quantum noise of: the EPR speedmeter with $\gamma_1=2\pi\times500\,{\rm Hz}$ and $\gamma_2=\sqrt{J/\gamma_1}\approx2\pi\times31\,{\rm Hz}$; the Michelson/Fabry-Perot position meter interferometer with $\gamma=2\pi\times500\,{\rm Hz}$; the Sagnac speed meter interferometer with $\gamma=(4J)^{1/3}\approx2\pi\times125\,{\rm Hz}$. In all cases, the phase quadrature of the output light is supposed to measured ($\zeta=\pi/2$).  All other parameters correspond to the ones of the planned LIGO Voyager GW detector, see Table \ref{tab:params}.}\label{fig:plots_S_1} 
\end{figure} 

Figures \ref{fig:plots_G} and \ref{fig:plots_S_1} show the optomechanical transfer function and the (single-sided) linear quantum noise spectral density, normalised to the equivalent GW signal:
\begin{equation}
  \sqrt{S^h} = \frac{2}{L}\sqrt{2S} \,,
\end{equation} 
respectively. For comparison, the corresponding plots for the Sagnac speed meter interferometer, also for the case of $\zeta=\pi/2$, and for a Michelson/Fabry-Perot (position meter) interferometer with the same value of the homodyne angle are also  provided.

It is easy to see from these plots that the EPR speed meter realises the trade-off between the high-frequency and the low-frequency sensitivity, losing discussed above 3\,dB to the position meter in shot noise dominated high- and medium-frequency region, yet providing significantly better sensitivity in the radiation-pressure noise dominated low-frequency region. It is also inferior by $\approx6\,{\rm db}$ to the Sagnac speed meter in the medium-frequency range. However, the relative simplicity of implementation of the EPR speed meter has the potential to easily outweigh this disadvantage.

\begin{figure}
  \includegraphics[scale=1.2]{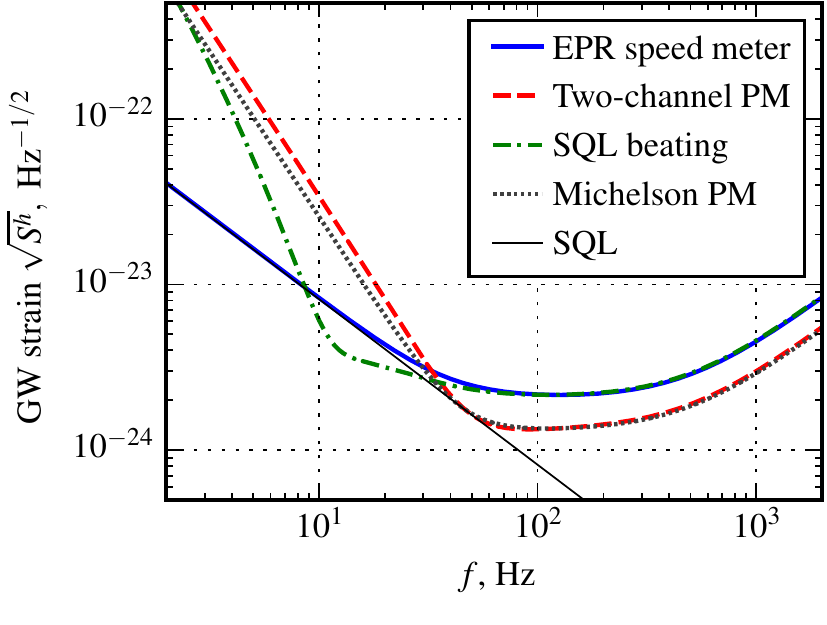}
  \caption{Quantum noise of the various regimes of the two-channel interferometer: $\zeta_+=0, \zeta_-=\pi/2$ (the EPR speedmeter); $\zeta_+=\zeta_-=\pi/2$ (the two-channel position meter); $\zeta_+=0.04, \zeta_-=1.4$ (an example of the SQL-beating tuning); the Michelson/Fabry-Perot position meter interferometer with $\zeta=\pi/2$. In all cases, $\gamma=\gamma_1=2\pi\times500\,{\rm Hz}$,  $\gamma_2=\sqrt{J/\gamma_1}\approx2\pi\times31\,{\rm Hz}$. All other parameters correspond to the ones of the planned LIGO Voyager GW detector, see Table \ref{tab:params}.}\label{fig:plots_S_2} 
\end{figure} 

An important feature of the of the EPR speed meter scheme is the flexibility provided by the two homodyne angles which can be tuned independently. In particular, if the phase quadratures are measured in both channels, $\zeta_+=\zeta_-=\pi/2$, then this schemes reduces to the position meter, see the corresponding plot in Fig.\,\ref{fig:plots_S_2}. The high-frequency sensitivity in this case is the same as in the ordinary Michelson positions meter. However, the low-frequency is worse by $\approx3\,{\rm db}$ due to the $\sqrt{2}$ times larger low-frequency asymptotic of the transfer function $\mathcal{G}_+$, in comparison with $\mathcal{G}_{\rm PM}$ (see Fig.\,\ref{fig:plots_G}), which increases proportionally the radiation-pressure noise.

The values of $0<\zeta_+,\zeta_-<\pi/2$ provide sensitivity better than the SQL. As Fig.\,\,\ref{fig:plots_S_2} shows, that (assuming the LIGO Voyager parameters) the SQL can be surpassed by a factor of up to $\sim3\,{\rm db}$ over a frequency range of half a decade. Note that the sub-SQL performance will increase with increasing power. The price for this is the sharp increase of the noise in the very low-frequency band due to incomplete cancellation of the ``$+$''-channel back action. Therefore, in any future application the EPR speed meter sensitivity can be optimised to obtain maximum reach for any astrophysical source for a given level of other noise sources of the GW detector.

\section{Discussion}

We would like to mention three important things here. First, the scheme described in this article is probably one of the simplest ways to realise a quantum speed meter, 
because it builds to a large extent on the traditional Michelson configuration with just minor extensions. Note that only a small number of additional optical elements will be needed, such as the additional quarter wave plate in the signal recycling cavity and the polarising beamsplitter at the output port. In addition it has to be ensured that the core interferometer can sustain normal operation for two orthogonal, linear polarisations.

Secondly, the proposed scheme can be switched back in-situ (using remote-controlled, in-vacuum waveplates) to the position meter mode, therefore increasing its medium- and high-frequency sensitivity. This transformation of the measurement mode can likely be obtained  within short time-scales and without any replacement of hardware, simply by switching the ``$+$'' output from the amplitude quadrature to the phase quadrature detection. If this switching and the corresponding changes to the control system and calibration system (e.g. adjusting loop gains and signal transfer functions) can be done fast enough and without introducing any additional displacement noise, then this possibility could be useful for  the  detection of GW signals from binary systems mergers, with the waveforms starting from very low frequency and gradually increasing to much higher ones.

Finally, it is evident that the capabilities of the considered scheme, featuring two outputs, are not limited by the simple configuration considered here. For instance the application of squeezed light states will even further improve the achievable strain sensitivity of our proposed scheme, and such options will be considered in in detail in a follow-up article.

\acknowledgements

This work was supported by the European Research Council Starter Grants program ERC-2012-StG:307245 and by the Royal Society International Exchanges Scheme Grant IE160125. The work of E.K. and F.K. was supported by Russian Foundation for Basic Research Grants 14-02-00399 and 16-52-10069. The work of F.K. was also supported by LIGO NSF Grant PHY-1305863.

The paper has been assigned LIGO document number LIGO-P1600351.

\appendix

\section{Post-processing of two output ports}

We assume that the photocurrents of the two homodyne detectors are combined using the optimal frequency-dependent weight function $\epsilon$:
\begin{equation}
  i_\epsilon(\Omega) = \epsilon(\Omega)i_+(\Omega ) + [1-\epsilon(\Omega)]i_-(\Omega) \,.
\end{equation} 
Note that this is a software operation and therefore no constrains are applied on the shape of $\epsilon$. This gives the following spectral density of the combined noise:
\begin{equation}\label{S_eps} 
  S_\epsilon = |\epsilon|^2 S_+ + 2\Re\bigl[\epsilon(1-\epsilon^*)S_\pm\bigr]
    + |1 - \epsilon|^2S_- \,,
\end{equation} 
where $S_+$, $S_-$ are the sum noise spectral densities at the ``$+$'' and the ``$-$'' outputs, respectively, and $S_\pm$ is the corresponding cross-correlation spectral density. It can be shown using Eqs.\,(\ref{io_pm}, \ref{F_fl_sum}), that they are equal to
\begin{subequations}
\begin{gather}
  S_+ = \frac{S_{\rm SQL}}{2}\left(
      \frac{1}{\mathcal{K}_+\sin^2\zeta_+} - 2\cot\zeta_+ + \mathcal{K}_+ + \mathcal{K}_-
    \right) , \\
  S_- = \frac{S_{\rm SQL}}{2}\left(
      \frac{1}{\mathcal{K}_-\sin^2\zeta_-} - 2\cot\zeta_- + \mathcal{K}_+ + \mathcal{K}_-
    \right) , \\
  S_\pm = \frac{S_{\rm SQL}}{2}
    (-\cot\zeta_- - \cot\zeta_+ + \mathcal{K}_+ + \mathcal{K}_-) ,
\end{gather}
\end{subequations}
where the factors $\mathcal{K}_+$, $\mathcal{K}_-$ are given by
\begin{equation}\label{K_p} 
  \mathcal{K}_+ = S_{\rm SQL}|\mathcal{G}_+|^2
  = \frac{[4\gamma_1^2\gamma_2^2 + \Omega^2(\gamma_1+\gamma_2)^2]J}
      {\Omega^2(\gamma_1+\gamma_2)(\gamma_1^2+\Omega^2)(\gamma_2^2+\Omega^2)} 
\end{equation} 
and by Eq.\,\eqref{K_EPR}.

The following optimization of \eqref{S_eps}:
\begin{equation}
  \epsilon 
  = -\frac{S_\pm^* - S_-}{S_+ + S_- - 2\Re S_\pm}
\end{equation} 
gives:
\begin{multline} \label{eq:OptimalNoise}
  S_{\rm opt} = \frac{S_- S_+ - |S_\pm|^2}
    {S_- + S_+ - 2\Re S_\pm} \\
  = \frac{S_{\rm SQL}}{2} %\\ \times
    \frac{
        1 - \mathcal{K}_+\sin2\zeta_+- \mathcal{K}_-\sin2\zeta_- 
        + \mathcal{K}_+^2\sin^2\zeta_+ + \mathcal{K}_-^2\sin^2\zeta_-
        + 2\mathcal{K}_+\mathcal{K}_-\sin\zeta_+\sin\zeta_-\cos(\zeta_+ - \zeta_-)
      }{\mathcal{K}_+\sin^2\zeta_+ + \mathcal{K}_-\sin^2\zeta_-} \,.  
\end{multline}

% \bibliographystyle{phaip}
% \bibliography{biblio_u,abbots_my,ligo,mqm}

\begin{thebibliography}{10}

\bibitem{CQG_32_7_074001_2015}
{J.Aasi {\it et al}},
\newblock Classical and Quantum Gravity {\bf 32}, 074001 (2015).

\bibitem{Affeldt_CQG_31_224002_2014}
C.~Affeldt et~al.,
\newblock Classical and Quantum Gravity {\bf 31}, 224002 (2014).

\bibitem{Accadia2012}
T.~Accadia et~al.,
\newblock Journal of Instrumentation {\bf 7}, P03012 (2012).

\bibitem{PRL_116_131102_2016}
{B.Abbott {\it et al}},
\newblock Phys. Rev. Lett. {\bf 116}, 131102 (2016).

\bibitem{PRL_116_241103_2016}
{B.Abbott {\it et al}},
\newblock Phys. Rev. Lett. {\bf 116}, 241103 (2016).

\bibitem{PRL_116_131103_2016}
{B.Abbott {\it et al}},
\newblock Phys. Rev. Lett. {\bf 116}, 131103 (2016).

\bibitem{Caves1981}
C.~M. Caves,
\newblock Phys. Rev. D {\bf 23}, 1693 (1981).

\bibitem{Nature_2011}
{J.Abadie {\it et al}},
\newblock Nature Physics {\bf 7}, 962 (2011).

\bibitem{Nature_2013}
{J.Aasi {\it et al}},
\newblock Nature Photonics {\bf 7}, 613 (2013).

\bibitem{92BookBrKh}
{V.B.Braginsky, F.Ya.Khalili},
\newblock {\em Quantum Measurement},
\newblock Cambridge University Press, 1992.

\bibitem{12a1DaKh}
{S.L.Danilishin, F.Ya.Khalili},
\newblock {Living Reviews in Relativity} {\bf 15} (2012).

\bibitem{90a1BrKh}
{V. B. Braginsky, F. Ja. Khalili},
\newblock Physics Letters A {\bf 147}, 251 (1990).

\bibitem{02a1KiLeMaThVy}
{H.J.Kimble, Yu.Levin, A.B.Matsko, K.S.Thorne and S.P.Vyatchanin},
\newblock Physical Review D {\bf 65}, 022002 (2001).

\bibitem{06a2Kh}
{F.Ya.Khalili},
\newblock Physical Review D {\bf 76}, 102002 (2007).

\bibitem{10a1Kh}
{F.Ya.Khalili},
\newblock Physical Review D {\bf 81}, 122002 (2010).

\bibitem{Evans_PRD_88_022002_2013}
M.~Evans, L.~Barsotti, P.~Kwee, J.~Harms, and H.~Miao,
\newblock Phys. Rev. D {\bf 88}, 022002 (2013).

\bibitem{00a1BrGoKhTh}
{V.B.Braginsky, M.L.Gorodetsky, F.Ya.Khalili and K.S.Thorne},
\newblock Physical Review D {\bf 61}, 044002 (2000).

\bibitem{Purdue2002}
P.~Purdue and Y.~Chen,
\newblock Phys. Rev. D {\bf 66}, 122004 (2002).

\bibitem{Chen2002}
{Y.Chen},
\newblock Physical Review D {\bf 67}, 122004 (2003).

\bibitem{04a1Da}
S.~L. Danilishin,
\newblock Phys. Rev. D {\bf 69}, 102003 (2004).

\bibitem{Graef_CQG_31_215009_2014}
C.~Gr\"af et~al.,
\newblock Classical and Quantum Gravity {\bf 31}, 215009 (2014).

\bibitem{Danilishin_NJP_17_4_043031_2015}
S.~L. Danilishin et~al.,
\newblock New Journal of Physics {\bf 17}, 043031 (2015).

\bibitem{Leavey_1603_07756}
{S.S.~Leavey {\it et al}},
\newblock arXiv:1603.07756  (2016).

\bibitem{Houston_speedmeter2017}
{E.A.Houston {\it et al}},
\newblock {\it paper in preparation}.

\bibitem{wp2016}
{LIGO Scientific Collaboration},
\newblock Instrument science white paper, 2016,
\newblock LIGO document T1600119,
  \url{https://dcc.ligo.org/LIGO-T1600119/public}.

\bibitem{00p1BrGoKhTh}
{V.B.Braginsky, M.L.Gorodetsky, F.Ya.Khalili and K.S.Thorne},
\newblock Energetic quantum limit in large-scale interferometers,
\newblock in {\em Gravitational waves. Third Edoardo Amaldi Conference,
  Pasadena, California 12-16 July}, edited by S.Meshkov, pages 180--189,
  Melville NY:AIP Conf. Proc. 523, 2000.

\bibitem{Caves1985}
C.~M. Caves and B.~L. Schumaker,
\newblock Phys. Rev. A {\bf 31}, 3068 (1985).

\bibitem{Schumaker1985}
B.~L. Schumaker and C.~M. Caves,
\newblock Phys. Rev. A {\bf 31}, 3093 (1985).

\bibitem{16a1DaKh}
F.~Y. Khalili and S.~L. Danilishin,
\newblock Progress in Optics {\bf 61}, 113 (2016).

\bibitem{Beyersdorf_JOSAB_16_1354_1999}
P.~T. Beyersdorf, M.~M. Fejer, and R.~L. Byer,
\newblock J. Opt. Soc. Am. B {\bf 16}, 1354 (1999).

\bibitem{Beyersdorf_OptLett_24_1112_1999}
P.~T. Beyersdorf, M.~M. Fejer, and R.~L. Byer,
\newblock Opt. Lett. {\bf 24}, 1112 (1999).

\bibitem{02a2Kh}
{F.Ya.Khalili},
\newblock arXiv:gr-gc/0211088  (2002).

\bibitem{Wade_PRD_86_062001_2012}
A.~R. Wade et~al.,
\newblock Phys. Rev. D {\bf 86}, 062001 (2012).

\bibitem{Wang_PRD_87_096008_2013}
M.~Wang et~al.,
\newblock Phys. Rev. D {\bf 87}, 096008 (2013).

\bibitem{Meers1988}
{B. J. Meers},
\newblock Physical Review D {\bf 38}, 2317 (1988).

\bibitem{Mizuno_PLA_175_273_1993}
J.~Mizuno et~al.,
\newblock Physics Letters A {\bf 175}, 273  (1993).

\end{thebibliography}

\end{document}